\documentclass[onecolumn,letterpaper,amsmath,amssymb,floatfix,aps,superscriptaddress]{revtex4}

\usepackage{graphicx}
\usepackage{dcolumn}  
\usepackage{bm}  
\usepackage{epsfig}

\begin{document}
\setlength\arraycolsep{2pt}

\title{Out of equilibrium thermal Casimir effect in a model polarizable material}

\author{David S. Dean}
\affiliation{Universit\'e de  Bordeaux and CNRS, Laboratoire Ondes et Mati\`ere d'Aquitaine (LOMA), UMR 5798, F-33400 Talence, France}
\affiliation{ Laboratoire de
Physique Th\'eorique (IRSAMC),Universit\'e de Toulouse, UPS and CNRS,  F-31062 Toulouse, France}
\author{Vincent D\'emery}
\affiliation{ Laboratoire de
Physique Th\'eorique (IRSAMC),Universit\'e de Toulouse, UPS and CNRS,  F-31062 Toulouse, France}
\author{V. Adrian Parsegian}
\affiliation{Department of Physics, University of Massachusetts, Amherst, MA, USA}
\author{Rudolf Podgornik}
\affiliation{Department of Theoretical Physics, J. Stefan
Institute, SI-1000 Ljubljana, Slovenia} 
\affiliation{Department of Physics, Faculty
of Mathematics and Physics, University of Ljubljana, SI-1000
Ljubljana, Slovenia}

\begin{abstract}
Relaxation of the thermal Casimir or van der Waals force for a model dielectric medium is investigated. We start with a model of interacting polarization fields  with a dynamics that leads to a frequency  dependent dielectric constant of  the Debye form. In the static limit the usual zero frequency Matsubara mode component of the Casimir force is recovered. We then consider the out of equilibrium relaxation of the van der Waals force to its equilibrium value when two initially uncorrelated dielectric bodies are brought into sudden proximity. It is found that the spatial dependence of the out of equilibrium force is the same as the equilibrium one but it has a time dependent amplitude, or Hamaker coefficient, which increases in time to its equilibrium value. The final relaxation to the equilibrium value is exponential in systems with a single or finite number of polarization field relaxation times. However, in systems, such as those described by the Havriliak-Negami dielectric constant, with a broad distribution of relaxation times, we observe a much slower power law decay to the equilibrium value.
\end{abstract}
\maketitle

\section{Introduction}
Despite the success of theoretical calculations on the equilibrium Casimir force \cite{Parsegian,Bordag,French}, aspects  of the out of equilibrium behavior are still poorly understood and the subject of controversy and debate. A number of approaches have been  adopted to compute thermal fluctuation induced forces out of equilibrium in simple models of soft matter systems and binary liquids . For example, the stress tensor has been used to compute the force \cite{ba2003,na2004,ga2006,ga2008,brito2011} in a variety of non-equilibrium contexts.
While it is clear that in such systems computations using the stress tensor will yield the
average value of the force at thermal equilibrium, it seems nevertheless that more information is needed regarding the dynamics of the field theory representing the critical or fluctuating field \cite{bitbol2011}, in particular how the value of the field at a surface changes when the surface is moved. An alternative approach is to define an energetic interaction of the field with a surface and then define forces via the 
principle of virtual work \cite{dego2009,dego2010}. Yet another is to define the force at a surface 
by a local kinetic arguments, for instance by using the ideal gas form for the pressure as a function
of the local density field \cite{br2007,bu2008}. There are notable differences in out of equilibrium 
forces computed using the approaches above and relatively few systems have been studied 
explicitly. However studies of free Gaussian field theories undergoing model A (non conserved)
dynamics have been carried out. Model A dynamics for the field is basically a diffusion equation 
driven by white noise and the thermal Casimir forces are found to tend toward the equilibrium
value with diffusive scaling \cite{ga2006,ga2008,dego2009,dego2010}.  As well as studying the
approach to equilibrium for dynamics obeying detailed balance, one can examine what happens when
the noise is non-thermal, for instance colored. In this case the steady state Casimir interaction, under 
model A type dynamics, tends to acquire an additional screening due to temporal correlations in the 
noise \cite{ba2003,dego2009,dego2010}.  We also note that the Parisi-Wu stochastic quantization scheme can be used to extract equilibrium results for the quantum Casimir force via a Langevin
dynamics approach \cite{rod2011}. 

In the case of quantum electrodynamics the definition of the instantaneous force can be derived
from the stress tensor as it is physically defined, within the theory of electromagnetism, via the 
force acting on charges and currents just as was done in the first Lifshitz paper on the equilibrium van der Waals force \cite{Lifshitz}. The study of the out of equilibrium quantum electromagnetic 
Casimir effect is however much more complicated than the equilibrium thermal Casimir effect and a number of physical assumptions need to be made if one wants to study
dynamics or non-equilibrium situations. One approach is to use generic models of fluctuating electrodynamics (or stochastic Maxwell equations) {\sl \` a la} Rytov \cite{Rytov},  where the field is driven by randomly fluctuating  current densities or dipole fields \cite{dor1998,pit2005,pit2008,Emig1,Emig2}. The out of equilibrium  context analyzed in the above studies is where the interacting media have different temperatures. In some cases there can be a long range bulk induced interaction between the media, essentially due to the  difference in their blackbody radiation. A similar bulk effect was found in \cite{na2004} for the  thermal Casimir effect in fluctuating scalar fields in the presence of temperature gradients.  While the theory of fluctuating electrodynamics is very general and applies to materials of 
generic dielectric properties,  there are nevertheless certain local equilibrium/fluctuation dissipation
properties that need to be assumed in these theories. The goal of this paper is not to study non-equilibrium  steady states but rather to address the question of how the equilibrium thermal Casimir force evolves in time to its equilibrium values, for instance when two dielectric slabs are brought suddenly into proximity.

Specifically we will examine the out of equilibrium behavior of the thermal part of the electrodynamic Casimir force (corresponding to thermal van der Waals interactions). Our approach will be based on a 
specialized form of fluctuating electrodynamics based on the full n-body dipolar interaction between two
model dielectric media.  The question we will address in this paper is how the thermal van der Waals interaction 
between two media relaxes to its equilibrium value?  Some physical properties of the thermal Casimir 
interaction look less mysterious when studied in this
setting and we will show that the thermal Casimir interaction is induced by the correlations between the
polarization fields of the two media. For a single type of polarization field our model corresponds to a Debye type
dielectric material. However, if we assume that there is a distribution of polarization fields with different
relaxation times and dielectric susceptibilities, then any dielectric function can be obtained by taking
a suitable distribution. Thus although our model only applies to Debye-like  dielectrics it can be applied
to any dielectric function as results of our computations only depend of the frequency dependent
dielectric function. Furthermore as we are interested in the approach to the equilibrium force we argue that the slowest relaxing polarizability fields will be the thermal ones (zero Matsubara frequency term) and among  these thermal modes those with the slowest dynamics should exhibit a Debye relaxation.  

The dynamics we assume for the  microscopic model for a polarizable dielectric media is over-damped 
stochastic dynamics for the polarization field. In the static limit, the force between such media corresponds 
to  the zero frequency Matsubara term in the  interaction energy between dielectric media as found in 
the Lifshitz theory \cite{Parsegian}.  We show here how the thermal van der Waals force between two objects emerges 
via the correlation of dipoles in the interacting media. Within this formalism we can see how the force 
evolves with time towards its equilibrium value, for instance, for two slabs brought into close proximity 
where initially they had infinite separation. We can also see how on upon changing the temperature 
of the system, the van der Waals force evolves from 
its initial equilibrium value at the initial temperature to the final equilibrium value at a different temperature. Although the model is simple, and restricted to the non-quantum part of the
van der Waals interaction it may give useful intuition about fluctuation induced forces out of equilibrium.
An appealing aspect of this approach is that the Laplace transform (with respect to time) of
the dynamical force can be expressed in terms of static results that can be obtained via well established 
equilibrium methods. 

\section{Model of polarizable media}

In this section we define a simple model for the polarizable media and their dielectric properties, showing that the dielectric response functions are a linear combination of Debye-like terms. We then  show how the average force between two such media in thermal equilibrium is identical to the thermal Casimir or van der Waals force  as predicted by the Lifshitz theory. 

Consider a system with an ensemble of  local polarization  fields ${\bf p}_\nu({\bf x})$ at the point ${\bf x}$ in the medium. The index $\nu$ corresponds to a type or species of polarization field which notably has its own  polarizability per unit volume denoted by   $\chi_\nu({\bf x})$. We consider two semi-infinite regions (slabs) $V^{+}$ and  $V^{-}$ defined via the sign of the coordinate $z$, such that $z>0$ in $V^+$ and $z<0$ in $V^-$. 
The two regions $V^+$ and $V^-$ will be separated in the $z$ direction by a distance $L$. In this 
notation the total energy for a given configuration of the dipole fields is  
\begin{equation}
H = {1\over 2} \int d{\bf x} d{\bf y}\sum_{\nu\nu'} {\bf p}_\nu({\bf x}) \cdot A_{\nu\nu'}({\bf x}, {\bf y},L ){\bf p}_{\nu'}({\bf y})
\label{equ1}
\end{equation}
where
\begin{equation}
A_{\nu\nu'}({\bf x},{\bf y}) = {\delta({\bf x}-{\bf y})I\delta_{\nu\nu'}\over \chi({\bf x})} + D({\bf x},{\bf y},L),
\end{equation}
In the first term  $I$ is the identity matrix in $3 D$ space and the polarization energy Eq. \ref{equ1} corresponds to the  classical harmonic energy needed to generate a local polarization field. In the second term  $D$ is the interaction between the dipoles in two semi-infinite regions which we will write in terms of the standard dipole-dipole coupling 
\begin{equation}
D_0({\bf x}-{\bf y}) = -\nabla_i\nabla_j G({\bf x}-{\bf y}),
\end{equation}
where $G$ is the Green's function for the electrostatic field {\sl in vacuo} obeying
\begin{equation}
\epsilon_0\nabla^2 G({\bf  x}) = -\delta({\bf x}).
\end{equation} 
Writing the energy in the above manner means that the separation ($L$) dependent part of the energy
is encoded in the dipolar interaction term $D$. This interaction is given by
\begin{equation}
D({\bf x},{\bf y},L) = D_0({\bf x}-{\bf y}),
\end{equation}
when ${\bf x}\in V^{+}$ and ${\bf y}\in V^{+}$ or when ${\bf x}\in V^{-}$ and ${\bf y}\in V^{-}$,
\begin{equation}
D({\bf x},{\bf y},L) = D_0({\bf x}-{\bf y }-L\hat{\bf z})
\end{equation}
when ${\bf x} \in V^{-}$ and ${\bf y} \in V^{+}$, and 
\begin{equation}
D({\bf x},{\bf y},L) = D_0({\bf x}+L\hat{\bf z}-{\bf y })
\end{equation}
when ${\bf x} \in V^{+}$ and ${\bf y} \in V^{-}$.  
 
Before proceeding with the analysis of Casimir interactions between two slabs, let us consider the 
dielectric properties of the bulk systems. We will assume that each dipole field obeys Langevin dynamics and that the only coupling between the different polarization fields are their mutual dipole-dipole interactions. The dynamical equations for each field in the presence of a uniform time dependent electric field ${\bf E}$ is thus  given by
\begin{equation}
{\partial {p_{i\nu}}({\bf x})\over \partial t} = -\kappa_\nu({\bf x}){\delta  H\over \delta {p_{i\nu}}({\bf x})} + \chi_\nu {E}_i(t)
+\zeta_{\nu i}({\bf x},t)
\end{equation}
where $\kappa_\nu$ is a local diffusion constant for each polarization field and determines the time-scale of relaxation of the field. The noise $\zeta_{\nu i}({\bf x},t)$ is of a white noise type and has a space-time correlation function
\begin{equation}
\langle \zeta_{\nu i}({\bf x},t) \zeta_{\nu' j}({\bf x}',t')\rangle = 2T \delta_{ij}\kappa_\nu({\bf x}) \delta_{\nu,\nu'}({\bf x},{\bf x}'),
\end{equation}
where $T$ is the temperature (imposed on the polarization field by the local bulk environment) and
the weighting $\kappa_\nu$ assures that detailed balance holds, {\em i.e.} that the dynamics will eventually
lead to thermal equilibrium at the temperature $T$.

In a bulk system (where $\kappa_\nu$ and $\chi_\nu$ are constant)  we determine the  dielectric properties of the 
model  by computing the response to a spatially uniform applied electric field ${\bf E}(t)= {\bf E}\exp(i\omega t)$. The average response of each polarizability field is uniform in space and the interactions between dipoles averages to zero. The average value of each polarizability field is then given by
\begin{equation}
\langle {\bf p}_\nu(t)\rangle = {\chi_\nu\over 1 + {i\omega \chi_\nu \over \kappa_\nu}}{\bf E}\exp(i\omega t),
\end{equation} 
and the average  total electric displacement is thus
\begin{equation}
{\bf D}(t) = \epsilon_0 {\bf E}(t) + \sum_{\nu}\langle {\bf p}_\nu (t)\rangle.
\end{equation}
This allows us to read off the frequency dependent dielectric response as 
\begin{equation}
\epsilon(\omega) =\epsilon_0 + \sum_\nu   {\chi_\nu\over 1 + {i\omega \chi_\nu \over \kappa_\nu}},
\label{fdec}
\end{equation}
which is obviously just a superposition of Debye-like dielectric responses. The frequency dependent
dielectric response thus depends on both the  polarizabilities $\chi_\nu$ and the dynamical
variable $\kappa_\nu$. The static dielectric constant however depends only on the static polarizabilities  
\begin{equation}
\epsilon(0) =\epsilon_0 + \sum_\nu   {\chi_\nu}. \label{statdec}
\end{equation}

\section{Equilibrium van der Waals interaction  for slab geometries}

After having established the basic bulk dielectric properties of our model we will
show how they enter the equilibrium van der Waals or thermal Casimir
effect between two semi-infinite dielectric slabs.

The force between two semi-infinite slabs separated by a distance $l$ for any configuration of the 
polarizability fields in the two slabs is given by
\begin{equation}
f= -{\partial H\over \partial L} = -{1\over 2}\sum_{\nu\nu'} \int d{\bf x} d{\bf y}\ {\bf p}_\nu({\bf x})\cdot {\partial \over \partial L}A({\bf x}, {\bf y},L ){\bf p}_{\nu'}({\bf y})
\label{equ2}
\end{equation}
since only the dipole interaction term $D$ depends on $L$. The local polarizability only depends on the coordinates within the two volumes $V^{+}$ and $V^{-}$. The equilibrium value for the
average force may be obtained using the fact that
\begin{equation}
\langle p_\nu({\bf x}) p_{\nu'}({\bf y})\rangle = {T}[A^{-1}({\bf x},{\bf y},L)]_{\nu\nu'}
\end{equation}
and thus  
\begin{equation}
\langle f\rangle =-{T\over 2} {\rm Tr}~\left(A^{-1} {\partial \over \partial L}A\right),
\end{equation}
where $\rm Tr$ indicates the trace over the operator, the spatial and the species indices. 
This average force may thus be written as
\begin{equation}
\langle f\rangle =-{T\over 2}  {\partial \over \partial L}{\rm Tr} \ln\left[A\right] = -{\partial F\over\partial  L},
\end{equation}
which of course agrees with the standard statistical mechanical definition if $F=-T\ln(Z)$ is interpreted as the free energy obtained from the partition function
\begin{equation}
Z = \int d[{\bf p}]\exp(-\beta H).
\end{equation}
This partition function can be written in a standard way by introducing a Hubbard-Stratonovich auxiliary field $\phi$, which physically can be identified with $i\psi$ where $\psi$ is the fluctuating electrostatic potential, to decouple the dipolar interaction. This gives up to a constant factor independent of $L$
\begin{eqnarray}
Z &=& \int \prod_{\nu}d[{\bf p}_\nu] d[\phi]\exp\left(-{\beta\over 2} \int {\epsilon_0}[\nabla \phi({\bf x})]^2 +\beta \int_{z<0}d{\bf x} \ i\sum_{\nu}{\bf p}({\bf x})\cdot \nabla \phi({\bf x}) -\sum_{\nu}{{\bf p}_\nu({\bf x})^2\over 2\chi_\nu({\bf x})} + \right. \nonumber \\&+& \left. \beta\int_{z>l} d{\bf x} \ i\sum_{\nu}{\bf p}_\nu({\bf x})\cdot \nabla \phi({\bf x}) -\sum_\nu{{\bf p}_\nu({\bf x})^2\over 2\chi_\nu({\bf x})}\right).
\end{eqnarray}
We note that the variables $\chi_\nu$ are not necessarily the same in the regions 
$z<0$ and $z>L$ as in general they will correspond to two different materials.
As the integrals over ${\bf p}_\nu$ are now decoupled, they may be carried out to yield, again up to constants independent of $L$, the standard form of the partition function for the thermal Casimir contribution for 
van der Waals interactions between dielectric media
\begin{equation}
Z = \int d[\phi] \exp\left(-{\beta \over 2} \int d{\bf x} \ \epsilon({\bf x},0) [\nabla \phi({\bf x})]^2\right).\label{pfs}
\end{equation}
Here $\epsilon({\bf x},0) = \epsilon_0 + \sum_\nu \chi({\bf x})$ for $z<0$ and $z>L$ {\em i.e.} in the volumes  $V^+$ and $V^-$ and $ \epsilon({\bf x}) = \epsilon_0$ for $z\in [0,L]$ {\em i.e.} the vacuum between the two media. We also see that the variables $\epsilon({\bf x},0)$ are simply the local static dielectric constants as defined by Eq. (\ref{statdec}). This form of the partition function corresponds to the one stemming from the Lifshitz theory of the Casimir force for the thermal  van der Waals
component of the interaction.

The partition function in Eq. (\ref{pfs})  can be evaluated exactly for systems where $\epsilon({\bf x})$ depends only on the coordinate $z$ (slab-like configurations), while for other geometries it can be evaluated via different approximation schemes such as the proximity force approximation or using systematic multipolar expansions. 

\section{Dynamics}

We now turn to the problem of the dynamical evolution of the force. In the absence of an applied field
the polarization dynamics in the two slabs can be written as 
\begin{equation}
{\partial {p_{i\nu}}({\bf x})\over \partial t} = -\kappa_\nu({\bf x}){\delta  H\over \delta {p_{i\nu}}({\bf x})} +
\zeta_i({\bf x},t)
\end{equation}
where $\kappa_\nu({\bf x})$ determines the local relaxation in the region ${\bf x}$ of space. 
The condition of detailed balance implies that the noise correlator obeys
\begin{equation}
\langle \zeta_{i\nu}({\bf x},t) \zeta_{j\nu'}({\bf x}',t')\rangle = 2T R_{i\nu\;  j\nu'}({\bf x},{\bf x}')
\end{equation}
where the operator $R$ is given by
\begin{equation}
R_{i\nu\; j\nu'}({\bf x},{\bf x}') = \kappa_\nu({\bf x})\delta_{ij}\delta_{\nu\nu'}\delta({\bf x}-{\bf x}').
\end{equation}
In operator notation the dynamical equations can be written as
\begin{equation}
{\partial {\bf p}({\bf x})\over \partial t}= -RA {\bf p}({\bf x}) + \zeta({\bf x},t).
\end{equation}
The average value of the dynamical force Eq. (\ref{equ2}) can be obtained from the time correlation function of the dipole field defined as
\begin{equation}
\langle p_{i\nu} ({\bf x},t) p_{j\nu'}({\bf x}',t)\rangle = C_{i\nu\; j\nu'}({\bf x},{\bf x}', t)
\end{equation}
so that the time dependent average force is given by
\begin{equation}
\langle f(t,L,T)\rangle =-{1\over 2} {\rm Tr}\ C(t){\partial \over \partial L}A.\label{fgen}
\end{equation}
The evolution equation for $\bf p$, being of first order in time, can be integrated to give an explicit form for the correlation
function 
\begin{equation}
C(t) = \exp(-tRA)C(0)\exp(-tA R) + T \Delta^{-1}(1-\exp(-2t A R))\label{eqC}
\end{equation}
in operator notation, where $C(0)$ is the value of the correlation function at $t=0$. We notice here that the force 
between the two regions depends on the cross correlation between them. If at $t=0$ the two regions are brought 
into proximity from a large distance the cross correlation at $t=0$ is zero and only the second term in  Eq. (\ref{eqC})
remains.  In the absence of initial correlations the Laplace transform of $C$
\begin{equation}
{\cal  L} C(s) =\int_0^\infty dt \ C(t)\exp(-st)
\end{equation}
is given by
\begin{equation}
{\cal L}C(s) = {T\over s}[A + {sR^{-1}\over 2}]^{-1}.
\end{equation}
This means that the Laplace transform of the time dependent average  force can be written as
\begin{equation}
{\cal L} \langle f\rangle (s) = -{T\over 2s} {\rm Tr}[A+ {sR^{-1}\over 2}]^{-1}{\partial \over \partial l}A.
\end{equation}
Two things should be noted at this stage: (i) the operator $R$ does not depend on the distance between the two regions  $V^+$ and $V^-$ and (ii) its inverse is simply
\begin{equation}
R^{-1}({\bf x},{\bf x}')_{\i\nu\; j\nu'} = {\delta_{ij}\delta_{\nu\nu'}\over \kappa_\nu({\bf x})}\delta({\bf x}-{\bf x}').
\end{equation}
The first of these points means that we can write
\begin{equation}
{\cal L} \langle f\rangle (s) = -{1\over s} {\partial F_s\over \partial L}={T\over s} {\partial \over \partial L}\ln(Z_s)
\end{equation}
where 
\begin{equation}
Z_s = \int \prod_{\nu} d[{\bf p}_\nu] \exp(-\beta H_d(s))
\end{equation}
and $H_d(s)$ is an effective dynamical Hamiltonian given by
\begin{equation}
H_d(s) = H + {s\over 4} \sum_\nu {{\bf p}_\nu^2({\bf x}) \over \kappa_\nu({\bf x}) }.
\end{equation}
It can be written as a static Hamiltonian with local dynamical polarizabilities
\begin{equation}
{1\over \chi_\nu (\bf x,s)} = {1\over \chi_\nu (\bf x)} + {s\over 2\kappa_{\nu}(\bf x)}
\end{equation} 
which in turn leads to  local {\em dynamic} dielectric constants given by
\begin{equation}
\epsilon_d({\bf x},s) = \epsilon_0 + \sum_\nu{\chi_\nu({\bf x})\over 1+ {s\chi_{\nu}({\bf x})\over 2\kappa_\nu({\bf x})}}.\label{eqed}
\end{equation}
The observant reader will immediately recognize a similarity between the form of Eq. (\ref{eqed}) for
the dynamical dielectric constant and the frequency dependent dielectric constant predicted by
the dielectric response model from Eq. (\ref{fdec}), indeed we find that $
\epsilon_d(s) = \epsilon\left(-i{s\over 2}\right)$.  Thus for the computation of the thermal van der Waals forces in this model, knowledge of the frequency dependent dielectric constants allows one to predict the temporal evolution of the force towards
its equilibrium value. Although the computation above was carried out with slab geometries in mind 
it is easy to see that it applies for general geometries where dielectric objects are separated by 
vacuum. This means that we can write the time dependent force as
\begin{equation}
\langle f(t) \rangle = -{\partial F(t, L)\over \partial L}
\end{equation}
where $F(t,L)$ is an effective time dependent free energy given by
\begin{equation}
F(t, L) = -T {\cal L}^{-1}{1\over s} \ln\left(Z(\epsilon(-i{s\over 2}))\right),
\end{equation}
where ${\cal L}^{-1}$ indicates the inverse Laplace transform and the notation $\epsilon(is/2)$ denotes that the frequency dependent dielectric constant is taken in all regions at the value $-is/2$. The pole at $s=0$ yields the equilibrium free energy and thus the equilibrium force. In addition we can invert the Laplace transform using the Bromwich integration formula to get
 \begin{equation}
F(t, L) = -T \int_{-i\infty}^{\infty} {ds\over 2\pi i\; s} \exp(st)\ln\left(Z(\epsilon(-i{s\over 2})\right).
\end{equation}
where the integration is to the right of the imaginary axis (as all singularities are at negative $s$). One can remove the term with the pole at $s=0$ by hand to write
\begin{equation}
F(t, L) = F_{eq}(L)  -T \int_{-i\infty}^{i\infty} {ds\over 2\pi i\; s} \exp(st)\left[\ln\left(Z(\epsilon(-i{s\over 2}))-
\ln(Z(\epsilon(0)\right)\right]
\end{equation}
where 
\begin{equation}
F_{eq}(L) = -T\ln\left(Z(\epsilon(0)\right)
\end{equation}
is the equilibrium free energy $F_{eq}(L)=\lim_{t\to\infty} F(t, L)$. In the above formula the remaining
contour integral is now free of singularities on the imaginary axis and we may therefore write it
as a Fourier transform by making the substitution $s=2i\omega$
\begin{equation}
F(t, L) = F_{eq}(L)  -T \int_{-\infty}^{\infty} {d\omega\over 2\pi i\; \omega} \exp(2i\omega t)[\ln(Z(\epsilon(\omega))-
\ln(Z(\epsilon(0))].
\end{equation}

The above  results can be generalized to the situation where the two slabs are in equilibrium
(at fixed distance $L$) at a temperature $T_0$ and the temperature is then changed to $T$. We can write the force for two slabs  as the sum of two components
\begin{equation}
\langle f(L,t, T_0\to T)\rangle = \langle f(L,T_0)\rangle_{eq} + {(T-T_0)\over T}\langle f^{(0)}(t,L,T)\rangle
\end{equation}
where $\langle f(L,T_0)\rangle_{eq}$ is the equilibrium force at the temperature $T_0$ and
$\langle f^{(0)}(L, t, T)\rangle$ is the time dependent force for two initially uncorrelated slabs at temperature $T$. 
The result for uncorrelated slabs at temperature $T$ can obviously be  extracted from the above result  by setting $T_0=0$. 

\section{Analytical  results for slab geometries}

Here we consider the case of two parallel semi-infinite slabs of dielectric constants $\epsilon_1$ and $\epsilon_2$ (we take the subscripts $1$ and $2$ to refer to the regions $V^-$ and $V^+$ respectively). The {\em dynamical free energy}
\begin{equation}
F_s = -T\ln\left(Z(\epsilon(-i{s\over 2})\right)
\end{equation}
can then be read off from standard equilibrium results and is given by
\begin{equation} 
F_s = {TS\over 16\pi L^2}\int_0^\infty u du\ \ln\left(1-\Delta_1(-i{s\over 2})\Delta_2(-i{s\over 2})\exp(-u)\right)
\end{equation} 
where $S$ is the area of the slabs and 
\begin{equation}
\Delta_i(\omega) = {\epsilon_{i}(\omega)-\epsilon_0\over \epsilon_i(\omega)+\epsilon_0}
\end{equation}
The time dependent force for two initially uncorrelated slabs is thus of the form
\begin{equation}
\langle f(L,t,T)\rangle = -{TS\over L^3}H(t)
\label{fundam}
\end{equation}
where $H(t)$ is a time dependent Hamaker coefficient whose Laplace transform is given by
\begin{equation}
{\cal L}H(s) =  -{1\over 8\pi s}\int_0^\infty u du\ \ln\left(1-\Delta_1(-i{s\over 2})\Delta_2(-i{s\over 2})\exp(-u)\right)
\end{equation}
and the superscript $0$ indicates that it is the force for initially uncorrelated slabs. The above expression for the time dependent Hamaker coefficient can then be written in terms of the polylogarithmic function
${\rm Li}_3(z) = \sum_{n=1}^\infty {z^n/n^3}$ to give
\begin{equation}
{\cal L}H(s) = {1\over 8 \pi s}{\rm Li}_3\left(\Delta_1(-i{s\over 2})\Delta_2(-i{s\over 2})\right).
\end{equation}
The static equilibrium value is simply recovered from the pole at $s=0$ outside
 and is given by
\begin{equation}
H_{eq} = {1\over 8 \pi}{\rm Li}_3\left(\Delta_1(0)\Delta_2(0)\right).
\end{equation} 
Eq. (\ref{fundam}) represents an interesting and fundamental result. It states that the non-equilibrium force has the same separation dependence as the equilibrium force, but its Hamaker coefficient is time dependent. At least on the non-retarded level the dynamic effects do not modify the spatial dependence of the van der Waals force and  the force is instantaneously long range. This decoupling of the spatial and temporal behavior of the force is in sharp contrast to that of critical  Casimir 
force for free  scalar fields with Dirichlet or Neumann boundary conditions undergoing model A dynamics. In that case the average force exhibits diffusive scaling, behaving  as $f(L,t)=t^{-\alpha}g(L/ \sqrt{t})$ \cite{ga2006,ga2008,dego2009,dego2010}.  We finally note that the mapping of the Laplace transform of the 
time dependent force onto an effective equilibrium problem is reminiscent of the results found in \cite{dego2009,dego2010} when a similar correspondence occurs for the time dependent thermal Casimir force for a free scalar field. 

Let us now investigate some special cases of the above general results.

\subsection{Short time behavior for Debye type dielectrics}

To begin with we consider the temporal evolution of the Hamaker coefficient  in the simplest case where the two bounding
surfaces are described by identical Debye type dielectric responses and the time evolution is limited to short time scales. Assuming that there is only one dipole type 
in each dielectric slab, we thus have
\begin{equation} 
\epsilon_i\left(-i{s\over 2}\right) = \epsilon_0 + {\chi_i\over 1+ {s \chi_i\over 2\kappa_i}}.
\end{equation} 
By dimensional analysis we see that we can write $\kappa_i = {\chi_i\over \tau_i}$, where $\tau_i$ is a microscopic polarization relaxation time in slab $i$. This change of notation
then yields the familiar Debye formula for the dielectric constant
\begin{equation}
\epsilon(\omega) = \epsilon_0 + {\Delta \epsilon\over 1 + i\omega \tau},
\end{equation} 
where  $\epsilon(0)- \epsilon_0 = \Delta\epsilon = \chi$. From this we obtain
\begin{equation}
\epsilon\left(-i{s\over 2}\right) = \epsilon_0 +{\epsilon_i-\epsilon_0\over 1 + {s\tau_i\over 2}},
\end{equation} 
which in turn yields
\begin{equation}
\Delta_i\left(-i{s\over 2}\right) = {\epsilon_i-\epsilon_0\over {\epsilon_i +\epsilon_0+\epsilon_0\tau_is}}.
\end{equation}
At short times the temporal behavior
of $H(t)$ can be obtained by looking at the large $s$ behavior of ${\cal L}H(s)$. In this limit we have that
\begin{equation}
\Delta_i\left(-i{s\over 2}\right)
\approx \left({\epsilon_i-\epsilon_0\over \epsilon_0}\right) {1\over s\tau_i}.
\end{equation}
As the  $\Delta_{i}\left(-i{s\over 2}\right)$ are small we are effectively in the dilute two body limit, this means that
the correlations other than two  body ones set in at later time scales. We thus find
\begin{equation}
{\cal L}H(s) \approx {(\epsilon_1-\epsilon_0)(\epsilon_2-\epsilon_0)\over 8\pi s^3 \tau_1\tau_2\epsilon_0^2}\label{stdeb}
\end{equation}
and thus inverting the Laplace transform we find that at short times
\begin{equation}
H(t)\approx {(\epsilon_1-\epsilon_0)(\epsilon_2-\epsilon_0)\ t^2\over 16\pi \tau_1\tau_2\epsilon_0^2}.
\end{equation}
The above initial growth is quadratic in time, reflecting the need for the polarization fields 
to become correlated. This result can be straightforwardly generalized to several polarization types $\nu_i$ in each
slab and we find that at early times
\begin{equation}
H(t)\approx {t^2\over 16\pi\epsilon_0^2}\sum_{\nu_1\ \nu_2} {\chi_{\nu_1}\chi_{\nu_2}\over \tau_{\nu_1}\tau_{\nu_2}}.\label{dst}
\end{equation}
From this expression one obviously discerns the pairwise nature of the interaction between the distinct polarization types
$\nu_1$ and $\nu_2$ in slabs 1 and 2. 

\subsection{Long time behavior for Debye type dielectrics}

Next we consider the long time behavior of the Hamaker coefficient  in the case of two identical 
Debye type materials. As we know the initial (and final equilibrium) values of the force we can
examine its full temporal evolution by analyzing the temporal derivative $\dot H(t) = dH/dt$. As the initial 
value of $H$ is zero, standard results on Laplace transforms give
\begin{equation}
{\cal L}\dot H(s) = {1\over 8 \pi }{\rm Li}_3\left(\Delta_1(-i{s\over 2})\Delta_2(-i{s\over 2})\right).
\end{equation}
In this case we have
\begin{equation}
\Delta_1(-i{s\over 2})=\Delta_2(-i{s\over 2})= {\epsilon-\epsilon_0\over \epsilon_0 \tau} {1\over s+a}
\qquad {\rm with} \qquad a = {\epsilon +\epsilon_0\over \epsilon_0 \tau}.
\end{equation}
Now using the series representation of ${\rm Li}_3$ we can invert the Laplace transform term by term 
to find
\begin{equation}
\dot H(t) = {1\over 8\pi}\sum_{n=1}^\infty \left({\epsilon-\epsilon_0\over \epsilon_0 \tau}\right)^{2n} {t^{2n-1}\exp(-at)
\over n^3 (2n-1)!} = {1\over 4\pi t} \exp(-at) R\left({(\epsilon-\epsilon_0)t\over \epsilon_0 \tau}\right),\label{re}
\end{equation}
where
\begin{equation}
R(u)= \sum_{n=1}^{\infty} {u^{2n}\over n^2 (2n)!} = 4\int_0^u ds\ \left(\ln(u) -\ln(s)\right) {\cosh(s)-1\over s} \simeq 2 {\exp(u)\over u^2} \qquad {\rm for} \qquad u \rightarrow \infty.
\end{equation}
The asymptotic form for $u \rightarrow \infty$ was derived by expressing $R(u)$ in terms of hypergeometric functions. Putting this all together yield for large $t$ 
\begin{equation}
\dot H(t) \approx {\tau^2\over 2\pi  t^3} \left({\epsilon_0\over \epsilon -\epsilon_0}\right)^2\exp(-{2t\over \tau}).
\end{equation}
Therefore at late times the asymptotic form of the time dependence of the Hamaker coefficient turns out to be 
\begin{equation}
H(t) \approx H_{eq} - {\tau^2\over 4\pi  t^2} \left({\epsilon_0\over \epsilon -\epsilon_0}\right)^2\exp(-{2t\over \tau}).
\label{debyeext}
\end{equation}
The apparent divergence in the second term above when $\epsilon\to\epsilon_0$ appears strange
at first sight, however we must bare in mind that the asymptotic expansion we carried out to 
obtain this result depended on the variable $u = (\epsilon-\epsilon_0)t/\tau\epsilon_0$ being large.
We can numerically verify the validity of the asymptotic expansion by comparison with  a direct
numerical evaluation of Eq. (\ref{re}). It is found to be correct but its realm of validity is for very large
values of $u$ of the order of $50$. The asymptotic expansion is thus of limited use and just shows that
at very late times the final relaxation to the equilibrium Hamaker coefficient is exponential with time scale
$\tau^* =\tau/2$, interestingly independent of $\epsilon$.  

A particularly non-trivial point about the above calculation is that though the pairwise approximation is valid for the equilibrium Hamaker coefficient when $\Delta \epsilon = \epsilon -\epsilon_0$ is small, the pairwise approximation cannot be used to extract the temporal behavior  of the out off equilibrium Hamaker coefficient. This is because $\Delta \epsilon$ appears multiplied by the time $t$ and thus the product of the two eventually must become large. Thus even when the  final equilibrium result is dominated by pairwise interactions, the dynamical evolution to the  equilibrium actually depends crucially on the full n-body interactions.

In the case where the two media are of Debye type but with different dielectric parameters the 
inversion of the Laplace transform of $\dot H(s)$ is more complicated and in general we have not been able to find an analytical expression as in the case where both slabs are composed of identical dielectric
media. However in  the case where the media are such that
\begin{equation}
a= a_1 = {\epsilon_1+\epsilon_0\over \tau_1 \epsilon_0}=a_2 = {\epsilon_2+\epsilon_0\over \tau_2\epsilon_0},
\end{equation}
this means that $\Delta_1(-i{s\over 2})$ and $\Delta_2(-i{s\over 2})$ have the same poles (at $s=-a_1 =-a_2$), we obtain
\begin{equation}
\dot H(t) = {1\over 4\pi t} \exp(-at) R\left({\sqrt{(\epsilon_1-\epsilon_0)(\epsilon_2-\epsilon_0)}t\over \epsilon_0 \sqrt{\tau_1\tau_2}}\right)
\end{equation}
Here the late time relaxation to the equilibrium Hamaker coefficient is again exponential with time
scale $\tau^* = 1/(a -\sqrt{(a-{2\over \tau_1})(a-{2\over \tau_2})})$, which we see is dependent of 
the dielectric constant but only through the variable $a$. 

A sub-variant of this situation, which can be analytically resolved, is the case where one of the systems has a much 
shorter relaxation time than the other, for instance $\tau_2 \ll \tau_1$. In the inversion of the 
Laplace transform we may use the approximation
\begin{equation}
\Delta_2\left(-i{s\over 2}\right) =  {\epsilon_2-\epsilon_0\over {\epsilon_2 +\epsilon_0+\epsilon_0\tau_2s}}
\approx {\epsilon_2-\epsilon_0\over {\epsilon_2 +\epsilon_0} }=\Delta_2(0),\label{aptau}
\end{equation}
{\em i.e.} this is essentially the assumption that the polarization field in the slab 2 instantaneously
equilibrates with the electric field produced by the polarization field of the slab. In addition the 
approximation is only valid for times $t$ such that $t\gg \tau_1$.
 Within this approximation we find that
\begin{equation}
\dot H(t) = {1\over 8\pi t}\exp(-a_1 t)W\left({\Delta_2 (\epsilon_1-\epsilon_0)t\over \epsilon_0 \tau_1}\right),
\end{equation}
where
\begin{equation}
W(u) = \sum_{n=1}^\infty {u^n\over n! n^2} \simeq {\exp(u)\over u^2} \qquad {\rm for} \ u \rightarrow \infty,
\end{equation}
in the large $u$ limit. Thus for large $t$ we remain with
\begin{equation}
H(t) \approx H_{eq} - {\tau_1^2\over 4\pi t^2} \left(\epsilon_0 \over \Delta_2 (\epsilon_1-\epsilon_0)\right)^2\exp\left( -{2 t (\epsilon_1 +\epsilon_2)\over \tau_1 (\epsilon_2 +\epsilon_0)}\right).
\end{equation}
Therefore even though the relaxation time of the slab 2 is very small, and the overall relaxation time scale 
is set by $\tau_1$, {\em i.e.} of the relaxation time of slab 1, the relaxation time of the full Hamaker coefficient scales 
with $\tau_1$ but depends also on the dielectric properties of slab 2 through the factor $(\epsilon_1 +\epsilon_2)\over (\epsilon_2 +\epsilon_0)$.  

\subsection{Havriliak-Negami type dielectrics}

It is possible to obtain non-Debye like behavior of the dielectric constant
by choosing a suitable distribution of  polarizability  and relaxation times for the associated
polarization fields, {\em i.e.} by assuming the existence of a distribution
\begin{equation}
\rho(\chi,\tau) = \sum_{\nu}\delta(\tau-\tau_\nu)\delta(\chi-\chi_\nu)
\end{equation}
such that
\begin{equation}
\epsilon(\omega)= \epsilon_0 +\int d\tau d\chi \ \rho(\chi,\tau)   {\chi\over 1+i{\omega \tau}}
\end{equation}
A very general phenomenological formula for the dielectric constant is the 
Havriliak-Negami formula \cite{hn1,hn2}
\begin{equation}
\epsilon(\omega) = \epsilon_0 +{\Delta\epsilon\over [1+(i\omega \tau_0)^{\alpha}]^\beta},
\end{equation}
with $\alpha \in[0,1]$ and $\beta >0$.
The Havriliak-Negami  form reduces to the Debye-model in the case where $\alpha=\beta=1$. When $\beta =1$ it gives the 
Cole-Cole formula and when $\alpha =1$ is gives the Cole-Davidson formula. We should note that the Havriliak-Negami dielectric function can be written explicitly as a superposition of individual
Debye relaxations and thus the study of this functional form within the dynamical formalism 
presented here is justified.  In terms of our model this model is composed of polarization fields 
of the same polarizability $\chi$ ($=\Delta \epsilon = \epsilon-\epsilon_0$) but with different relaxation times $\tau$. This
means that the dielectric function can be written in the form
\begin{equation}
\epsilon(\omega)= \epsilon_0 +\Delta\epsilon\int d\tau  \ \rho(\tau)   {\chi\over 1+i{\omega \tau}},
\end{equation}
where 
\begin{equation}
\rho(\tau) = {1\over \tau \pi} {({\tau\over \tau_0})^{\alpha\beta}\sin(\beta\theta)\over
\left[ ({\tau\over \tau_0})^{2\alpha} + 2 ({\tau\over \tau_0})^\alpha +1\right]^{\beta\over 2}} \qquad {\rm and} \qquad \theta = \tan^{-1}\left[ {\sin(\pi\alpha)\over ({\tau\over \tau_0})^{\alpha} + \cos(\pi\alpha)}\right].\label{disnn}
\end{equation}
In the short time limit (corresponding to large $s$) we find
\begin{equation}
{\cal L}H(s) \approx {2^{\alpha_1\beta_1+\alpha_2\beta_2 -5}\Delta\epsilon_1 \Delta\epsilon_2\over \pi s^{1+\alpha_1\beta_1 +\alpha_2\beta_2} \tau_{01}^{\alpha_1\beta_1}
\tau_{02}^{\alpha_2\beta_2}\epsilon_0^2},
\end{equation}
which after inverting the Laplace transform gives for short times
\begin{equation}
H(t) \approx
{2^{\alpha_1\beta_1+\alpha_2\beta_2 -5}\  t^{\alpha_1\beta_1 +\alpha_2\beta_2} \Delta\epsilon_1
\Delta\epsilon_2\over \pi  
\Gamma(\alpha_1\beta_2+\alpha_2\beta_2 +1)\tau_{01}^{\alpha_1\beta_1}
\tau_{02}^{\alpha_2\beta_2}\epsilon_0^2}\label{negst}
\end{equation}
where $\Gamma(z)$ is the Euler gamma function. We thus see that the exponents $\alpha$ and $\beta$ in the Havriliak-Negami formula control the early time growth exponent which depends on the product of the two $\alpha\beta$. We can also verify that this general formula agrees with Eq. (\ref{dst}) in the Debye case where all $\alpha$ and $\beta$ are equal to one. 

The late time decay to the equilibrium Hamaker coefficient between two slabs with Havriliak-Negami dielectric functions can be extracted from the small $s$ expansion of ${\cal L}H(s)$ and we find
\begin{equation}
H(t) \approx H_{eq} - {\epsilon_0\over \pi} \rm{Li}_2(\Delta_1(0)\Delta_2(0)) \left( {\beta_1\over (\epsilon_1+ \epsilon_0)\Gamma(-1-\alpha_1)}\left({\tau_{01}\over 2t}\right)^{\alpha_1}  + {\beta_2\over (\epsilon_2+ \epsilon_0)\Gamma(-1-\alpha_2)}\left({\tau_{02}\over 2t}\right)^{\alpha_2}\right).\label{hnlt}
\end{equation} 
The late time decay is thus dominated by the term above with the smaller value of $\alpha_i$. This form of the Hamaker coefficient relaxation is very different from the pure Debye exponential relaxation, Eq. (\ref{debyeext}), and obviously shows a long time algebraic tail. We thus see that the time needed to relax  to equilibrium in this case can be much longer than the microscopic time scales $\tau_{0i}$ that set the characteristic time of Hamaker coefficient relaxation in the pure Debye model. 

\section{Numerical results}
 
In this section we will numerically compute the time evolution of the Hamaker coefficient to its 
equilibrium value by numerically inverting the Laplace transform $H(s)$. 
\begin{figure}
\begin{center}
\resizebox{0.6\hsize}{!}{\includegraphics[angle=0]{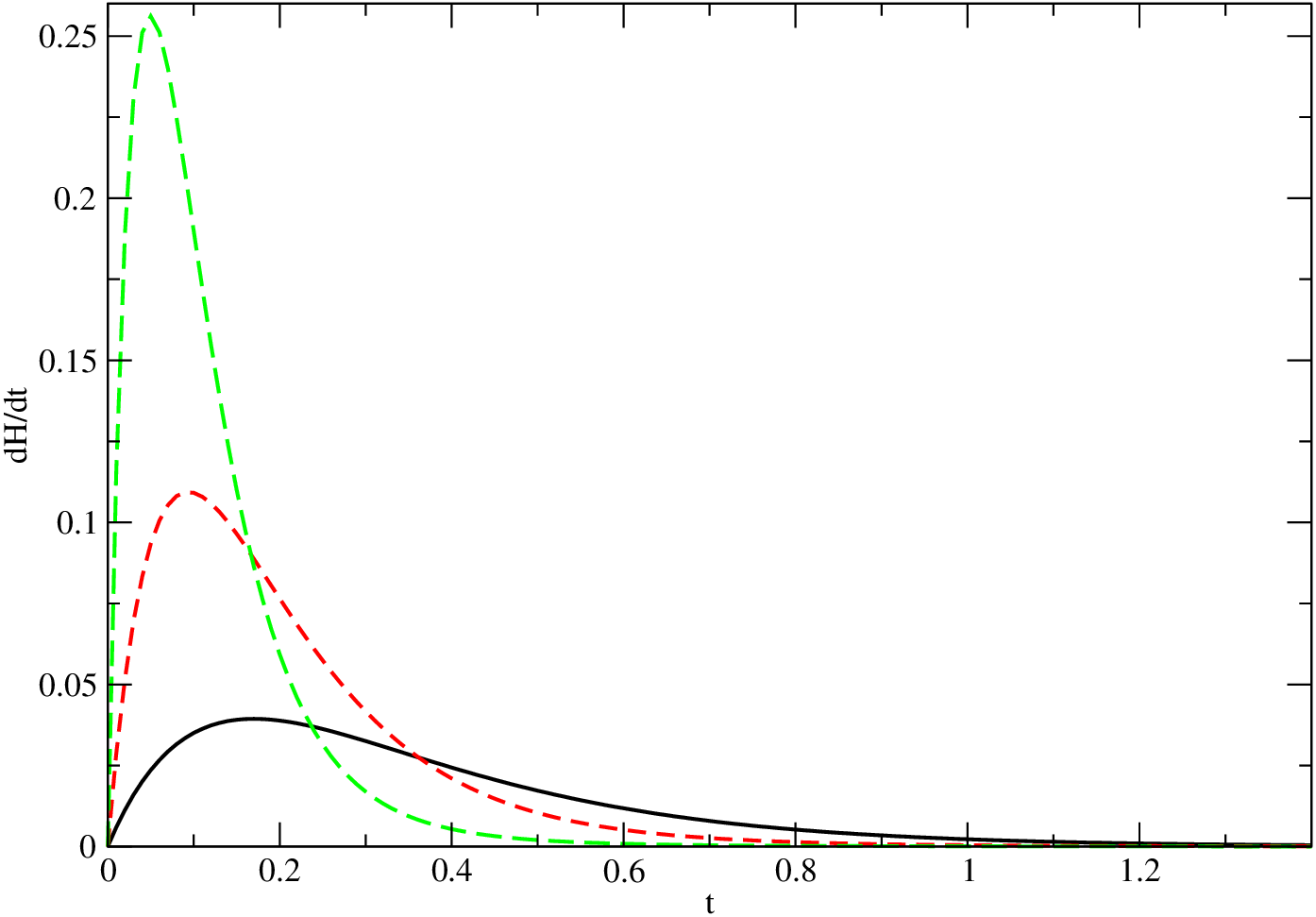}}
\end{center}
\caption{Color online. Time derivative  of the Hamaker constant $\dot H(t)$ for  initially uncorrelated
Debye slabs made up from identical media all with with $\tau = 1$ and 
(i) $\epsilon = 5 \epsilon_0$ (solid black line) (ii) $\epsilon = 5 \epsilon_0$ (short dashed red line) (iii)
$\epsilon = 20 \epsilon_0$ (long dashed green line)}
\label{debanal}
\end{figure}
\subsection{Debye dielectrics}

The most complete analytical results we have obtained are for the temporal evolution of the Hamaker constant for two Debye dielectrics composed of the same material. We have a closed form expression for the temporal derivative $\dot H(t)$  given by Eq. (\ref{re}). In Fig. (\ref{debanal}) we have compared 
the analytical form of Eq. (\ref{re}) with the result obtained by  numerical inversion of the Laplace transform for $\dot H(t)$ for the cases $\tau=1$ and $\epsilon = 5,\ 10$ and $20$. Only the analytical curves are shown as the accord with the analytical formula is perfect. The figure shows that the temporal
derivative increases mots rapidly at short times for the systems of higher dielectric constant as predicted by Eq. (\ref{stdeb}). However the final relaxation is slowest for the systems of lower dielectric constant.

Now we consider a pure Debye case where each slab is characterized by a single 
time scale $\tau_i$ and dielectric constant $\epsilon_i$. We examine the
case where $\epsilon_1 =10\epsilon_0$ and $\epsilon_2=2\epsilon_0$ in the following three cases where  (i) $\tau_1=1$, $\tau_2=1$ (ii) $\tau_1=1$, $\tau_2=.1$ and (iii) $\tau_1=.1$ $\tau_2=1$ 
(therefore we are measuring time in the units of the larger of the times $\tau_1$ and $\tau_2$. Note
that the equilibrium $H_{eq}$ value of the Hamaker coefficient is the same in all of these cases . In Fig ({\ref{debye}) we plot the functions $H(t)/H_{eq}$ for each cases. We see that in the last two cases, where the shorter time scale $\tau =0.1$ is introduced, the approach to the equilibrium value is 
quicker. However the quickest relaxation occurs when the shorter of the two relaxation times is associated with the more dilute dielectric medium, {\em i.e} that with the lower dielectric constant.
\begin{figure}
 \begin{center}
\resizebox{0.6\hsize}{!}{\includegraphics[angle=0]{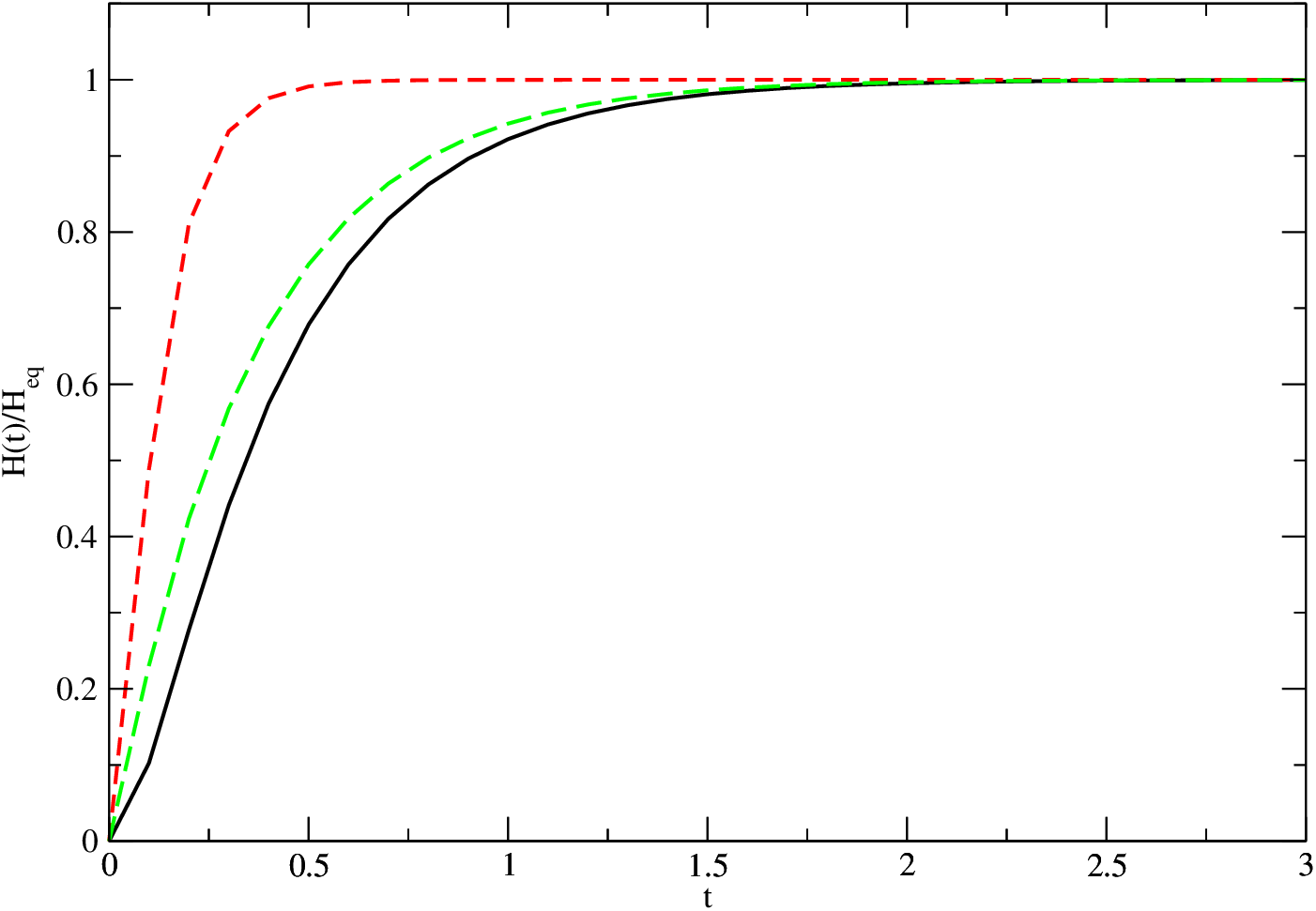}}
\end{center}
\caption{Color online. Time evolution of Hamaker constant normalized by its equilibrium value for two initially uncorrelated slabs pure Debye-like dielectric slabs 1 and 2 with $\epsilon_1 = 10\epsilon_0$, $\epsilon_2 = 2 \epsilon_0$ and relaxation times (i) $\tau_1=1$, $\tau_2 =1$ (solid black line)
(i) $\tau_1=1$, $\tau_2 =0.1$ (short dashed red line) (iii) (i) $\tau_1=0.1$, $\tau_2 =1$ (long dashed green line)}
\label{debye}
\end{figure}

\subsection{Havriliak-Negami  dielectrics}

We now consider the Nevriliak-Negami form for the dielectric constant of two medium. For simplicity we
consider media of the same type with a fixed $\Delta \epsilon = \epsilon-\epsilon_0$ and we will 
use units such that $\tau_0 = 1$. The case where $\Delta \epsilon = 3$ and when $\beta = 1$ (the Cole-Cole) case is shown in Fig. (\ref{cole}) for several values of $\alpha$. We see that as $\alpha$ decreases from 1 toward zero, the relaxation to the final equilibrium Hamaker coefficient
(which is the same for all the curves as the static dielectric constants are the same) becomes 
increasingly slow as one would expect from Eq. (\ref{hnlt}). Also, numerical fitting of the late time decay toward 1 in Fig (\ref{cole}) is compatible with the analytic prediction of the late time exponents given
in Eq. (\ref{hnlt}). The initial behavior is the inverse, the systems with smallest
$\alpha$ have a $H(t)$ which grows faster, in accordance with the predictions of Eq. (\ref{negst}).
An interesting feature of Fig (\ref{cole}) is that all the curves cross each other at the same,
isosbestic-like, point in time,  at around $t= 0.25$. 

\begin{figure}
 \begin{center}
\resizebox{0.6\hsize}{!}{\includegraphics[angle=0]{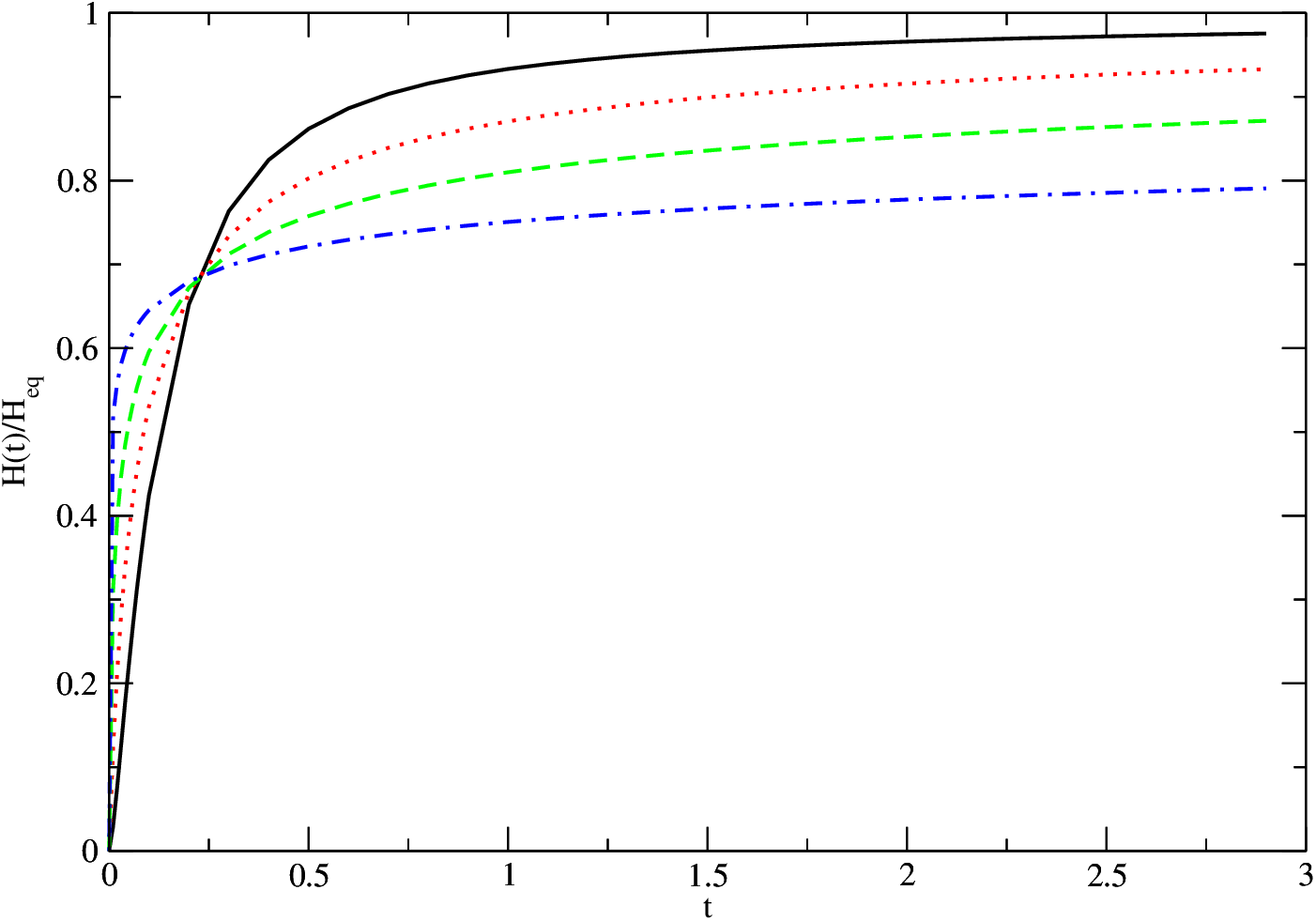}}
\end{center}
\caption{Color online. Time evolution of Hamaker consant normalized by its equilibrium value for two identical initially uncorrelated  Cole-Cole (Nevriliak-Negami with $\beta =1$) dielectric slabs 1 and 2 with $\epsilon = 10\epsilon_0$,  and characteristic relaxation times  $\tau=1$ for different values of $\alpha$:
 (i) $\alpha=0.8$ (solid black line) (ii) $\alpha = 0.6$ (red short dashed) (iii) $\alpha =0.4$ (green long dashed) (iv) $\alpha = 0.2$ (blue dot-dashed).}
\label{cole}
\end{figure}

In Fig (\ref{coled}) we  show the temporal evolution of $H(t)$ for Cole-Davidson type dielectrics for
different values of $\beta$ but with the same dielectric constants. As predicted from Eq. (\ref{negst})
the systems with the smaller values of $\beta$ show the most rapid growth in $H(t)$ at short times but
in contrast with the Cole-Cole case these systems also converge most rapidly to the equilibrium value, and there is no isosbestic-like point in this case.

\section{Discussion}

We have formulated a theory describing how the thermal van der Waals force between two dielectric objects evolves with time towards its equilibrium value. As an example of a general approach we analyze the dynamics of the thermal Casimir or zero Matsubara frequency van der Waals interactions between two slabs brought into close proximity  from an infinite separation, corresponding to the limit of the non-quantum part of the total van der Waals  interaction. Despite these simplifications our calculations give a useful intuition about fluctuation induced forces out of equilibrium. The particular strength of our approach is that  the Laplace transform (with respect to time) of the dynamical force can be expressed in terms of equilibrium results available from a wide range of  equilibrium methods. 

We show that for the zero frequency Matsubara term in planar geometry the coarse-grained dynamics of the interacting dielectric media enters only via the Hamaker coefficient, while the dependence on the spacing between the dielectric  interfaces remains unchanged and coincides with the equilibrium scaling. This is a fundamental result but is  limited to non-retarded form of the interaction only. It is however this part of van der Waals interaction that is most important in (bio)colloid- and nano-systems.

The time evolution of the non-retarded van der Waals force between two surfaces of area $S$ at temperature $T$ separated by $L$, $f(L, t, T)$, is then given by 
\begin{equation}
\langle f(L, t ,T)\rangle = -{TS\over L^3}H(t).
\end{equation}
While the separation into an equilibrium separation scaling and a non-equilibrium Hamaker coefficient appears to be universal for non-retarded interactions, the form of the time dependence of the Hamaker coefficient, $H(t)$, is specific and pertains to the dielectric response model of the two bounding dielectric surfaces. 

We have shown that for Debye-type dielectric response of the interacting materials with a relaxation time of $\tau$, the scaling of the Hamaker coefficient for short times is given by $$H(t) \sim C_{<} ~t^2,$$while in the asymptotic time regime we obtain $$H(t) \approx H_{eq} - {C_{>}\over t^2}  \exp(-{2t\over \tau}).$$The time scale of these non-equilibrium effects in thermal Casimir interactions is thus determined by the (longest) dielectric relaxation time of the interacting media. This would render the practical observation of these non-equilibrium effects difficult in general, however it may be possible to observe temporal evolution of the force  for systems with  extremely long relaxation times such as polymers and colloids and glassy systems. 

Indeed in  the case of the  non-Debye-like response that we have studied , i.e. the Havriliak-Negami dielectric response function (which is commonly applied to polymeric systems), we obtain a completely different asymptotic behavior of the Hamaker coefficient. Instead of an exponential scaling with the relaxation time, a  long algebraic tail is obtained instead. We derived the scaling form $$H(t) \approx H_{eq} - \left(C_{>} \over 2t\right)^{\alpha}$$ in the late time asymptotic regime, where $\alpha$ is one of the scaling exponents in the Havriliak-Negami dielectric response function. This long time algebraic tail in the relaxation of the non-equilibrium Hamaker coefficient leads to the conclusion that for this particular dielectric model it might be possible to observe long time non-equilibrium effects.

Variation in the form of the time evolution of the non-equilibrium Hamaker coefficient with the nature of the dielectric response of the interacting media makes it possible, or indeed quite probable, that in some experiments where at least indirectly a time-dependent 
van der Waals interactions are probed, these effects may complicate a clear cut interpretation of the experiments. This would be especially true for the tapping mode AFM measurements of macromolecular interactions or any other situation involving time varying separation between the interacting dielectric interfaces. The AFM tapping mode vibration of the interacting surfaces together with the time evolution of the appropriate non-equilibrium Hamaker coefficients would lead to a non-trivial modification 
of the dependence of the magnitude of the van der Waals interaction on the dielectric characteristics of the interacting surfaces.  Interestingly enough, at least for the non-retarded component of the interaction, the spatial dependence should remain intact.
The non-equilibrium effects considered in this paper would thus complicate the quantitative interpretation of experiments on van der Waals interactions in terms of the dielectric susceptibilities of the interacting interfaces.  In order to interpret these experiments  one would thus possibly need to incorporate the theory presented above to extract quantitative measures of time-averaged van 
der Waals interactions.

Another possible area of further analysis would be a comparison between these dynamical non-equilibrium van der Waals forces with the hydrodynamic drag forces of the Stefan type. These forces have their origin in the flow of fluid from between the two interacting 
surfaces as they are pushed together \cite{Russell}. The time dependence of the non-equilibrium van der Waals interaction, and the specific model studied here,  could also
be related to, and shed light on, the phenomenon of drag induced on moving
bodies by Casimir forces \cite{ann1986,mkr,pen2010,dem2011}.

\begin{figure}
 \begin{center}
\resizebox{0.6\hsize}{!}{\includegraphics[angle=0]{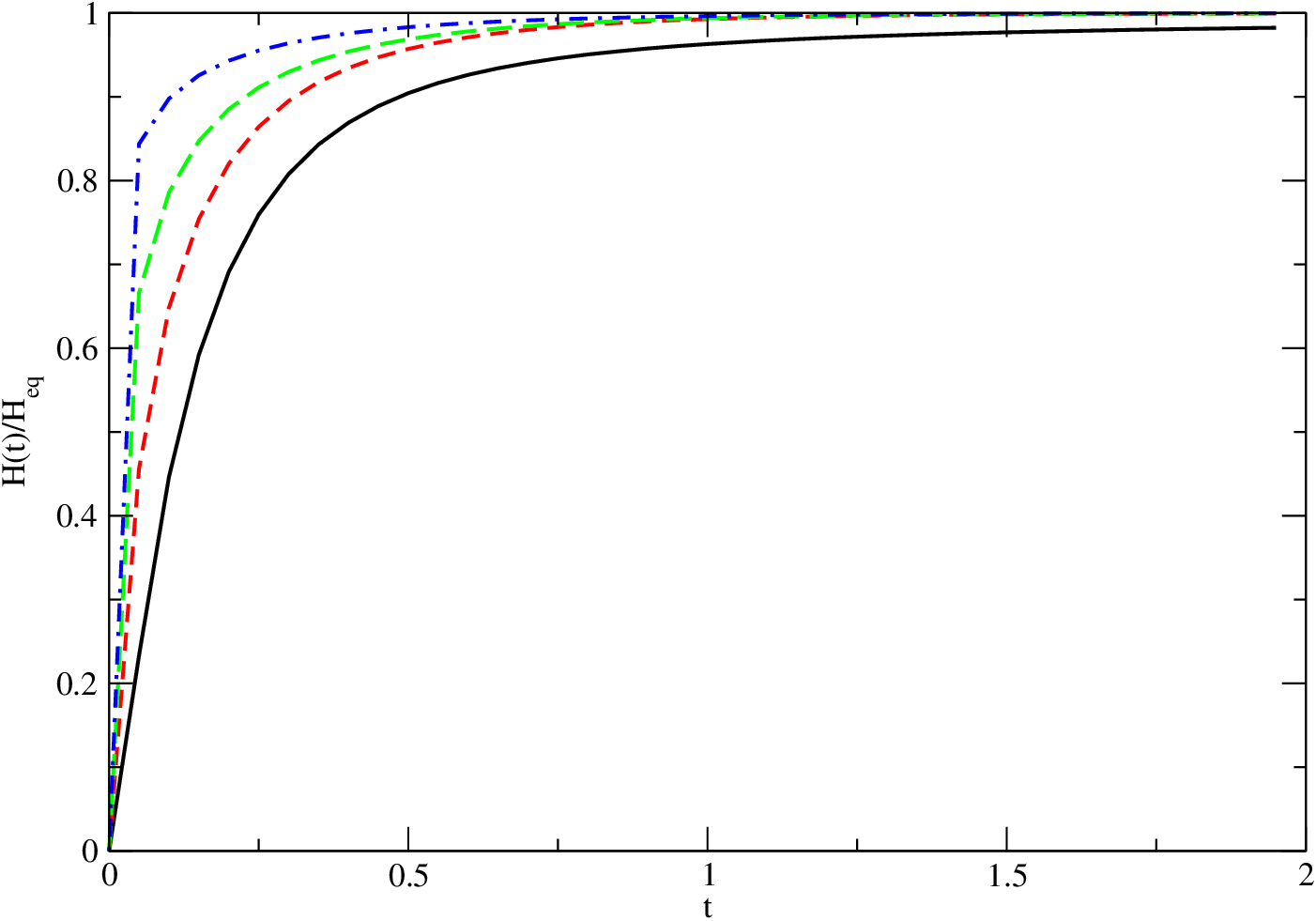}}
\end{center}
\caption{Color online. Time evolution of Hamaker consant normalized by its equilibrium value for two identical initially uncorrelated  Cole-Davidson (Nevriliak-Negami with $\alpha =1$) dielectric slabs 1 and 2 with $\epsilon = 10\epsilon_0$,  and characteristic relaxation times  $\tau=1$ for different values of $\alpha$:
 (i) $\beta=0.8$ (solid black line) (ii) $\beta = 0.6$ (red short dashed) (iii) $\beta =0.4$ (green long dashed) (iv) $\beta = 0.2$ (blue dot-dashed).}
\label{coled}
\end{figure}


\end{document}